\documentclass[sigconf]{acmart}

\usepackage[utf8]{inputenc}

\AtBeginDocument{%
  \providecommand\BibTeX{{%
    \normalfont B\kern-0.5em{\scshape i\kern-0.25em b}\kern-0.8em\TeX}}}

\copyrightyear{2019}
\acmYear{2019}
\acmConference[CUI 2019]{1st International Conference on Conversational User Interfaces}{August 22--23, 2019}{Dublin, Ireland}
\acmBooktitle{1st International Conference on Conversational User Interfaces (CUI 2019), August 22--23, 2019, Dublin, Ireland}
\acmPrice{15.00}
\acmDOI{10.1145/3342775.3342809}
\acmISBN{978-1-4503-7187-2/19/08}

\begin{document}

\title{A Need for Trust in Conversational Interface Research}

\author{Justin Edwards}
\email{justin.edwards@ucd.ie}
\affiliation{%
  \institution{University College Dublin}
  \country{Ireland}
}

\author{Elaheh Sanoubari}
\email{e.sanoubari@cs.umanitoba.ca}
\affiliation{%
  \institution{University of Manitoba}
  \country{Canada}
}

\renewcommand{\shortauthors}{Edwards and Sanoubari}

\begin{abstract}
  Across several branches of conversational interaction research including interactions with social robots, embodied agents, and conversational assistants, users have identified trust as a critical part of those interactions. Nevertheless, there is little agreement on what trust means within these sort of interactions or how trust can be measured. In this paper, we explore some of the dimensions of trust as it has been understood in previous work and we outline some of the ways trust has been measured in the hopes of furthering discussion of the concept across the field.
\end{abstract}

\ccsdesc[500]{Human-centered computing~HCI theory, concepts and models}
\ccsdesc[500]{Human-centered computing~Interaction design theory, concepts and paradigms}
\ccsdesc[300]{Human-centered computing~User centered design}

\keywords{trust, design, conversational agents, Human-robot interaction, Human-agent interaction}


\maketitle

\section{Introduction}
Trust is an important dimension in how people use any technology and development or extinction of it plays a large role in the ultimate adoption of the technology \cite{young_toward_2009}. Trust is specifically important in scenarios where people need to accept information provided by an agent and follow its suggestions to benefit from it \cite{freedy_measurement_2007}. Designing such interfaces with trust in mind can help maintain effective relationship with agents. 

One example of these interfaces is conversational user interfaces (CUIs) which can include text-based dialogue systems, voice-based conversational assistants, embodied virtual agents, and social robots. While CUIs have become increasingly prevalent in the last several years, numerous user studies have revealed that people have concerns about trusting these interfaces \cite{clark_what_2019, olson_2019_2019, cowan_what_2017}. Nevertheless, there has been little discussion in the community on how to define trust in these interactions, how to measure trust, or how designers can make interfaces more trustworthy \cite{torre_measuring_2018}.

Users can have a range of issues with CUIs that fall under the umbrella of trust, yet have fundamental differences. For example, a person may have mistrust in a robot because they think it is faulty and not computationally capable of doing a task \cite{salem_would_2015}, or they may not trust it because they think it is malicious and trying deceiving them \cite{short_no_2010}. We argue that it is important to differentiate different aspects of trust and what shapes them as only by understanding and operationalizing trust will it be possible to meet user needs when designing CUIs. In this paper, we outline some of the possible dimensions to consider when defining trust for CUIs and explore some avenues for measuring trust in conversational interactions.

\section{Defining Trust}
The first step in designing for trust is understanding and operationalizing what is meant when CUI users mention the term. While other topics in human-computer interaction (HCI) have attempted to map the meaning and role of trust in their field (e.g. automation \cite{lee2004trust} and e-commerce \cite{egger2000trust}) this  has not yet occurred in CUI interactions. As interactions with these interfaces are frequently modeled on human interactions both by designers and implicitly by users \cite{cowan_what_2017}, there may be reason to believe that trust in this context has a social or relational purpose. This conceptualization of trust, what has been operationalized as a willingness to become vulnerable to others in the social sciences \cite{rousseau_not_1998}, has been applied in HCI previously in the context of information exchanges on websites \cite{mcknight_impact_2002}. 

CUI user studies have revealed different meanings of trust however. One study of intelligent personal assistant (IPA) users found many users referring to issues around trust in the CUI’s intents, for example regarding identity and privacy protections \cite{clark_what_2019}. Here, users were not concerned with disclosure of sensitive information because of a social uneasiness, but instead because they were unsure how their data would be treated and what data was being collected in the first place. This notion of trust has been previously explored in the context of media consumers level of trust in media sources \cite{kiousis_public_2001} and may help to frame this version of trust in HCI. 

Other user studies have revealed a conceptualization of trust that relates instead to a belief in the abilities and competency of the interface \cite{luger_like_2016}. This definition of trust more closely resembles the concept of credibility or believability which has been well studied in the context of websites and news media credibility \cite{flanagin_role_2007, fogg_elements_1999}. 

Furthermore, trust or lack of it can be analyzed by where it is originated and how it is shaped. That is, in human-computer interaction trust can be synthesized by traits in both the computer and the human. For example, different people can have different attitudes towards the similar technologies which can be shaped by factor such as cultural differences \cite{haring_perception_2014}, personality traits \cite{salem_would_2015}, context/task \cite{wang_when_2010}, familiarity \cite{de_visser_almost_2016}, etc. 

Clearly, these definitions of trust are heterogeneous and the single concept of trust may have multifaceted meanings for people interacting with CUIs. A better understanding of these dimensions of trust is needed as CUI research moves forward.

\section{Measuring trust} 
Because of the multifaceted nature of trust, it is clear that a mixed-methods approach would be necessary in measuring different dimensions of the construct. Some of these methods might include behavioural measures, subjective questionnaires, analysis of body language and linguistic choices, and physiological measures. 

Previous assessments of trust have taken varied approaches according to the way the concept was operationalized. This sort of operationalizing of trust has been done in automation research to better capture the concept through emprically supported factors \cite{hoff2015trust}. Much work in interactions literature has been done on the Big Five personality traits \cite{john_big_1999}, in which trustworthiness is considered a component of agreeableness and thus a stable aspect of personality. Work looking at trust this way has used Likert-scale questionnaires to gauge people’s ratings of those traits in embodied conversational agents \cite{celiktutan_automatic_2017}. 

Other work has viewed trust as a trait that is acquired through behavioural experience and thusly measured trust though people’s choices in an economic game like the classic prisoner’s dilemma \cite{torre_trust_2018}. Additionally, measures of interpersonal trust in human-human dialogues and measures of functional kinds of trust for e-commerce systems have been correlated or been proposed to be studied alongside physiological measures like heart rate \cite{mitkidis_building_2015} and eye movements \cite{riegelsberger_shiny_2003}. Methods involving laboratory experiments are of course costly to utilize, so analysis of existing or easily collected data are likewise important. In some studies of trust as an interpersonal construct in human-human dialogues, trust has been measured through corpus analysis of linguistic features \cite{scissors_cmc_2009} and nonverbal cues \cite{lee_computationally_2013}. Analyses like these may be particularly useful in quickly measuring trust for conversational agents deployed over the web like commercial IPAs and chatbots. Combining subjective, behavioral, and physiological measures of trust not only enrich the mixed methods approach proposed here for studying trust, but they may also, in combination, allow for creation of robust validated measures which are lacking in conversational HCI in general \cite{clark_state_2018}.

\section{Conclusion}
Trust has been recognized as an integral dimension in how people request and use information from machines, how people physically interact with machines, and how people work alongside machines. While trust has been explored in different areas of HCI and social sciences in previous work, the growing research areas that involve CUIs need definitions and measurements for understanding the construct in conversational contexts. Because these interactions are fundamentally modeled on human-human interactions and applied to computers, they must be informed by research both in social science and HCI. Further design of CUIs requires our community to engage deeply with social constructs like trust in order to better understand user behaviour and facilitate adoption of these technologies.

\begin{acks}
This research was supported by Science Foundation Ireland (SFI) award number: 13/RC/2106 ADAPT. The authors would like to thank Ilaria Torre, Leigh Clark, Benjamin Cowan, and all attendees of Measuring and Designing Trust in Human-Agent Interaction at HAI 2018 for valuable discussion of this topic.
\end{acks}

\bibliographystyle{ACM-Reference-Format}
\bibliography{trust}

\end{document}